\documentclass[%
twocolumn, prl, aps, superscriptaddress, longbibliography, showpacs, amsmath, amssymb, floatfix
]{revtex4-1}

\usepackage{graphicx}
\usepackage{dcolumn}
\usepackage{bm}
\usepackage{graphicx}                                               
\usepackage{amssymb}

\usepackage{amsmath}
\usepackage{epsfig}
\usepackage{xcolor}
\usepackage{tabu}
\usepackage{mathtools}
\usepackage[colorlinks,linkcolor=blue,anchorcolor=blue,citecolor=blue,urlcolor=blue]{hyperref}
\usepackage{physics}
\usepackage{float}
\usepackage{diagbox}
\usepackage{inputenc}

\begin{document}

\title{From single-particle to many-body mobility edges and the fate of overlapped spectra in coupled disorder models}
\author{Xiaoshui Lin}
\affiliation{CAS Key Laboratory of Quantum Information, University of Science and Technology of China, Hefei, 230026, China}
\author{Ming Gong}
\email{gongm@ustc.edu.cn}
\affiliation{CAS Key Laboratory of Quantum Information, University of Science and Technology of China, Hefei, 230026, China}
\affiliation{Synergetic Innovation Center of Quantum Information and Quantum Physics, University of Science and Technology of China, Hefei, Anhui 230026, China}
\affiliation{Hefei National Laboratory, University of Science and Technology of China, Hefei 230088, China}
\author{Guang-Can Guo}
\affiliation{CAS Key Laboratory of Quantum Information, University of Science and Technology of China, Hefei, 230026, China}
\affiliation{Synergetic Innovation Center of Quantum Information and Quantum Physics, University of Science and Technology of China, Hefei, Anhui 230026, China}
\affiliation{Hefei National Laboratory, University of Science and Technology of China, Hefei 230088, China}

\date{\today}


\begin{abstract}
Mobility edge (ME) has played an essential role in disordered models. However, while this concept has been well established in disordered single-particle models, its existence in disordered many-body models is still under controversy. 
Here, a general approach based on coupling between extended and localized states in their overlapped spectra for ME is presented. 
We show that in the one-dimensional (1d) disordered single-particle models, all states are localized by direct coupling between them.
However, in $d \ge 2$ disordered single-particle and 1d disordered many-body models, the resonant hybridization between these states in their overlapped spectra makes all states be extended, while these in the un-overlapped spectra are unchanged, leading to tunable MEs.
We propose several models, including two disordered many-body spin models, to verify this mechanism.
Our results establish a unified mechanism for MEs and demonstrate its universality in single-particle and many-body models, which opens an intriguing avenue for the realization and verification of MEs in many-body localization. 
\end{abstract}

\maketitle


Mobility edges (MEs), the energy separating localized states and extended states in Anderson localization (AL), have attracted lots of attention in disordered Anderson models \citep{anderson_absence_1958, abrahams_scaling_1979, MacKinnon1981OneParameter, Belitz1994Anderson, Lagendijk2009Fifty}, quasiperiodic models \citep{Biddle2009Localization, Biddle2010Predicted, Ganeshan2015Nearest,  Wang2020one-dimensional}, and even non-linear disorder models \citep{WangObservation2022, Alex2021Interaction, Skipetrov2008Anderson}, etc. 
Recently, this concept has even been extended to many-body localization (MBL) models \citep{Pal2010Many-body, Nandkishore2015Many-body, Abanin2019Manybody, Alet2018Many-body} and was identified as the energy separating the ergodic states from the MBL states.
However, while the MBL can be viewed as some kind of AL in the Hilbert space with correlated disorders \citep{Tikhonov2021Anderson}, the fate of MEs in AL and MBL is completely different. 
Namely, while the single-particle ME in AL \citep{Pasek2017Anderson, Luschen2018Single-particle, Kohlert2019Observation,  Semeghini2015Measurement, Bai2021Learning} has been well-established, the existence of many-body ME is still controversial \citep{Deroeck2016absence, Deroeck2017Stability, Wei2019Investigating, Luitz2015Many-body, Brighi2020Stability, Deng2017Many-body, Chanda2020Many-body, Nag2017many-body, Modak2015Many-body,  Zhang2022Localization, Lazarides2015Fate}, due to the different structures of the Hilbert space involved and the different ways the extended and localized states interact in AL and MBL models \citep{Deroeck2016absence, Luitz2017SmallBath, Thiery2018Delocalization, Leonard2023Probing}.
Thus, despite the intimate similarities between these two systems, the connection between AL and MBL is still unclear, which hinders the construction of many-body MEs in the interacting models and their experimental verification \cite{Kohlert2019Observation, Luschen2018Single-particle}. 

\begin{figure}[htbp]
\centering
\includegraphics[width=0.35\textwidth]{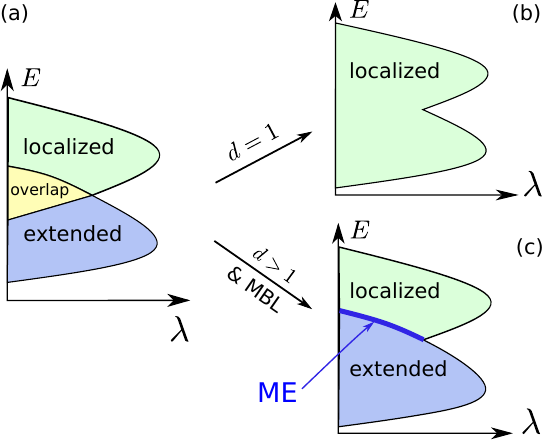}
\caption{(a) Extended and localized states without direct coupling. (b) In 1d disordered single-particle models, all states are localized due to the coupling between them. 
(c) In 2d and 3d disordered single-particle and 1d many-body models, the states in the overlapped regime will become extended by their direct coupling, while the un-overlapped spectra are unchanged, leading to ME.}
\label{fig-schematic-me}
\end{figure}

Here, we demonstrate the tunable MEs using coupled disordered models with two components belonging to two different phases. 
We find the following significant conclusions:
(I) In the overlapped spectra, the coupling between localized and delocalized states may lead to fully localized states in one-dimensional (1d) models and fully extended states in higher-dimensional models.
In the un-overlapped spectra, the properties of the wave functions in higher-dimensional models are unchanged, leading to MEs.  
(II) The above mechanism applies equally well to disordered single-particle and many-body models, in which the 1d disordered many-body models can be regarded as some kind of higher dimensional localization in the Hilbert space. 
We have constructed several experimentally feasible models to support the above conclusions. 
Consequently, the concept of MEs in single-particle and many-body models is unified into the same picture. 
This kind of random model with ME, itself, is also of fundamental importance in random matrices \cite{Forrester2010LogGas}, which has not yet been fully understood. 

\textit{The coupled disorder model for ME}: This model can be abstractively described by the following blocked matrix
\begin{equation}
    \mathcal{H} = \begin{pmatrix}
        H_\text{ext} & H_{c} \\
        H_{c}^{\dagger} & H_\text{loc}
    \end{pmatrix},
    \label{eq-couple-matrix}
\end{equation}
where $H_\text{ext}$ ($H_\text{loc}$) represents the Hamiltonian of the extended (localized) component and $H_\text{c}$ is the inter-component coupling.
When $H_\text{c} = 0$, Eq. \ref{eq-couple-matrix} is block diagonal and the spectra of $H_\text{ext}$ and $H_\text{loc}$ are independent, which may be overlapped in some regimes; see Fig. \ref{fig-schematic-me} (a).
Then we switch on $H_\text{c}$, which is still weak, the coupling between the two blocks $H_\text{loc}$ and $H_\text{ext}$ can lead to new physics in the overlapped regime.
In the one-dimensional (1d) models, the random disordered potential localizes all states; see Fig. \ref{fig-schematic-me} (b).
However, in $d\geq 2$ single-particle and  1d many-body models, $H_c$ will lead all states in the overlapped spectra to be extended from their hybridization, following Mott's argument \citep{Mott1967Electrons, Luca2011Localized}, yet the properties of the un-overlapped spectra are unchanged (see Fig. \ref{fig-schematic-me} (c)), yielding ME.
This idea can also be generalized to random matrices, which will be discussed elsewhere \cite{Lin2023Model}. 

\textit{Numerical methods}: Several complementary approaches are employed to characterize the extended and localized states.
(1) For small systems, we use the exact diagonalization method to obtain all eigenstates and eigenvalues.
For large systems ($N \gg 10^4$), we use the shift-invert method \citep{Pietracaprina2018Shift, Luitz2015Many-body} to obtain several ($N_E = 6-20$) eigenstates and eigenvalues around a given energy $E$; see S1 \citep{SI}.
Then we characterize the localized and extended phases using the consecutive level-spacing ratio \citep{Oganesyan2007Localization, Atas2013Distribution}
\begin{equation}
r_n = \min(s_n,s_{n+1})/\max(s_n,s_{n+1}),
\end{equation} 
where $s_n = E_n - E_{n-1}$, with $E_n$ being the eigenvalues in ascending order, yielding $\langle r\rangle_{\text{GOE}} = 0.5307(1)$ for extended states 
(Gaussian orthogonal ensemble, GOE) and $\langle r\rangle_{\text{PE}} = \ln(4) - 1 \sim 0.3863$ for AL (Poisson ensemble, PE) \citep{Tarquini2017Critical, Liu2022Localization} and MBL \citep{Pal2010Many-body, Iyer2013Many-body, Cheng2021Many-body}, respectively. 
The distribution of $r_n$ is denoted by $P(r)$ \cite{SI}.
(2) We identify the localization of the eigenstate $\psi$ using the inverse participation ratio (IPR), $\langle \mathrm{IPR}\rangle_E =  \langle \sum_m|\psi_{m}|^4 \rangle_E$, from which we have the fractal dimension as \cite{Evers2008Anderson}
\begin{equation}
    \tau_2(E,L) = -\ln( \langle \text{IPR}\rangle_E )/\ln(L),
\end{equation}
with $\langle \cdot \rangle_E$ denoting its averaged value in a narrow interval of $E$ under $10^2 \sim 10^3$ disorder realizations.
We also define $\tau_2(E) = \lim_{L \rightarrow \infty} \tau_2(E, L)$ \cite{SI}, and have $\tau_2(E) = 0$ for the localized phase; and $\tau_2(E) = d$ ($d$ is the spatial dimension) for the extended phase \citep{Evers2008Anderson}. 
(3) We use the transfer matrix method to obtain the Lyapunov exponent $\gamma(E)$ \citep{SI, Hoffmanbook}.
To analyze the finite-size effect, we define the dimensionless Lyapunov exponent $\tilde{\gamma}(E) = M \gamma(E) $ for higher dimensional systems with $M$ the length of one cross-section.
It shows $\tilde{\gamma}(E) \sim M/\xi$ for the localized states and $\tilde{\gamma}(E) \sim (\xi/M)^{d-2}$ for the extended states \citep{Hoffmanbook, Tarquini2017Critical}.
(4) Finally, we calculate the normalized half-chain entanglement entropy $S_\text{EE}/L = -\text{tr}(\rho_{A}\ln(\rho_A)) $ for the disordered many-body models with $\rho_A$ the reduced density matrix of the left half-chain. 
This quantity can distinguish the volume law behavior ($S_\text{EE}/L \sim $ finite) of the ergodic states from the area law behavior ($S_\text{EE}/L \sim 0$) of the MBL states \citep{Luitz2015Many-body, Abanin2019Colloquium}. 

\begin{figure}[htbp]
\centering
\includegraphics[width=0.48\textwidth]{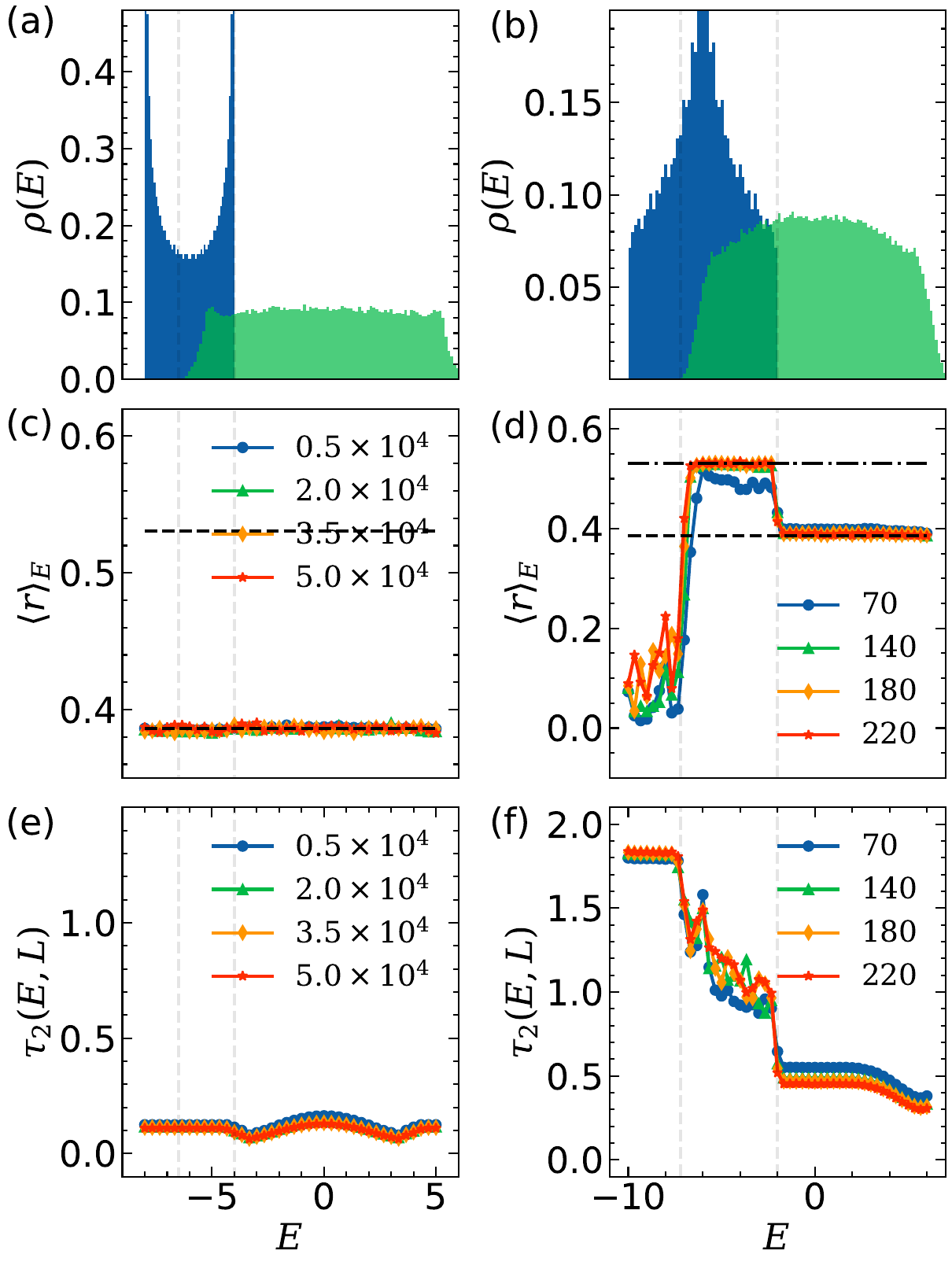}
\caption{
(a), (b) The average density of states, $\rho(E)$, of the uncoupled ($t_\text{v} = 0$) model, in which the blue (green) shadow represents $\rho(E)$ in $H_\text{ext}$ ($H_\text{loc}$).
(c), (e) The level-spacing ratio $\langle r\rangle_E$, and fractal dimension $\tau_2(E, L)$ for 1d model with sizes $L = (0.5 - 5.0) \times 10^4$. (d), (f) The corresponding results in 2d model with $L = 70 - 220$ ($N=L^2$).
The vertical grey dashed lines denote the boundaries of the overlapped spectra. Here we have used $V_1=M_2=0$, $V_2=10$, $M_1=-6$, $t_\text{v}=0.1$ in all figures.}
\label{fig-1d-2d-al}
\end{figure}

\textit{(I) AL in 1d coupled model and ME in 2d coupled model}: 
Dimension is one of the key factors in AL \citep{abrahams_scaling_1979}.
It is well-known that in 1d disordered models, all states should be localized.
Yet AL transition can happen in 1d models with incommensurate potential \citep{aubry1980analyticity, Biddle2010Predicted, Wang2020one-dimensional}.
The two-dimensional (2d) disordered models are marginal.
While all states can be localized in the spinless disordered models, extended states with ME can be realized in some 
spin-orbit coupled models \cite{Hikami1980Spin, Xie1998Kosterlitz, Wang2015Band, Chen2019Metal}. To this end, we consider the following spinful Hamiltonian 
\begin{equation}
H = \sum_{\langle i,j\rangle,\alpha} t a_{i\alpha}^{\dagger}a_{j\alpha} + \sum_{i\alpha} W_{i\alpha} a_{i\alpha}^{\dagger}a_{i\alpha} + t_\text{v} a_{i\alpha}^{\dagger} a_{i\bar{\alpha}},
\label{eq-al-model}
\end{equation}
with $\alpha$ ($\bar{\alpha}$$) = \uparrow, \downarrow$ ($\downarrow, \uparrow$) is the component index, $t =1$, and  $\langle i,j\rangle$ represents the nearest neighbor hopping.
The on-site potential $W_{i,\alpha} \sim U[M_{\alpha}-V_{\alpha}/2, M_{\alpha}+V_{\alpha}/2]$ is a uniformly distributed random number, with offset controlled by $M_\alpha$. 
By changing $M_\alpha$, the overlapped regime is changed, which yields tunable ME for $d\geq 2$ models (see S2 of Ref. \citep{SI}).
We set $L$ as the length of the system in one direction, thus the total number of sites $N = 2L^d$, with $d$ being the system dimension.
The model in Eq. \ref{eq-al-model} can be directly realized in ultracold atoms with a spin-dependent random potential \citep{Yang2017Spin-dependent, Mandel2003Coherent, Darkwah2022Probing}. 
When $t_\text{v} = 0$, It represents two copies of Anderson models, in which states in $H_\text{ext}$ (with $V_1 = 0$) are fully extended, and in $H_\text{loc}$ are localized when $V_2 \ne 0$ 
(when $d=1, 2$). In the presence of $t_\text{v}$, the fate of the overlapped spectra will be changed dramatically, following the picture of Fig. \ref{fig-schematic-me} (b) and (c).

We first present our results for the 1d model in Fig. \ref{fig-1d-2d-al} (a) and (c), where local mean level-spacing ratio $\langle r\rangle_E \sim 0.39$ for the localized phase. 
The fractal dimension is also very small with $\tau_2(E, L) < 0.2$, which decreases slightly with the increase of system size $L$.
In S3 \citep{SI}, we perform the finite-size scaling of $\tau_2(E, L)$ and find it approaches zero for all $E$ when $L\rightarrow\infty$.
Thus all states are localized in the presence of inter-component coupling.
This result is consistent with the scaling argument in Ref. \cite{Abrahams1979Scaling}

MEs in higher dimensional models are allowed. 
The results for the 2d coupled model are shown in Fig. \ref{fig-1d-2d-al} (b) and (d). 
We find that in the overlapped spectra regime, the local $\langle r\rangle_E \sim 0.53$ and $\tau_2(E, L) \rightarrow 2$ with increasing $L$, indicating extended phase.
We have also found $\langle r\rangle_E \sim 0.386$ and $\tau_2(E, L) \rightarrow 0$ for states with $E>-2$,
which correspond to the localized states.
For the states with $E<-7.4$, $\tau_2(E, L) \rightarrow 2$, indicating that these states are extended.
However, in this regime, $\langle r\rangle_E \sim 0 - 0.2$ are very small due to eigenvalue degeneracy in the clear Hamiltonian, which still exhibits strong level-spacing repulsion with increasing lattice size ( see S4 \citep{SI}).
The finite-size scaling of $\tau_2(E, L)$ up to $L=1450$ is presented in S3 \citep{SI}, which also support this conclusion.
These extended states in the 2d model do not contradict with the scaling argument \citep{Abrahams1979Scaling}, in which AL in 2d is marginal.
In fact, it has been known for a long time that spin-orbit coupling, which plays the same role as inter-component coupling, can lead to delocalization in the 2d system \cite{Hikami1980Spin, Chen2019Metal}, with even random magnetic fields \citep{Xie1998Kosterlitz, Wang2015Band}.
Similar delocalization is also reported in the 2d quantum Hall systems \citep{Kagalovsky1997Landau-level, Avishai2002New, Xiong2001Metallic, Evers2008Anderson}. 
Our model presents a more practical way for MEs in 2d. 

\begin{figure}[htbp]
\centering
\includegraphics[width=0.48\textwidth]{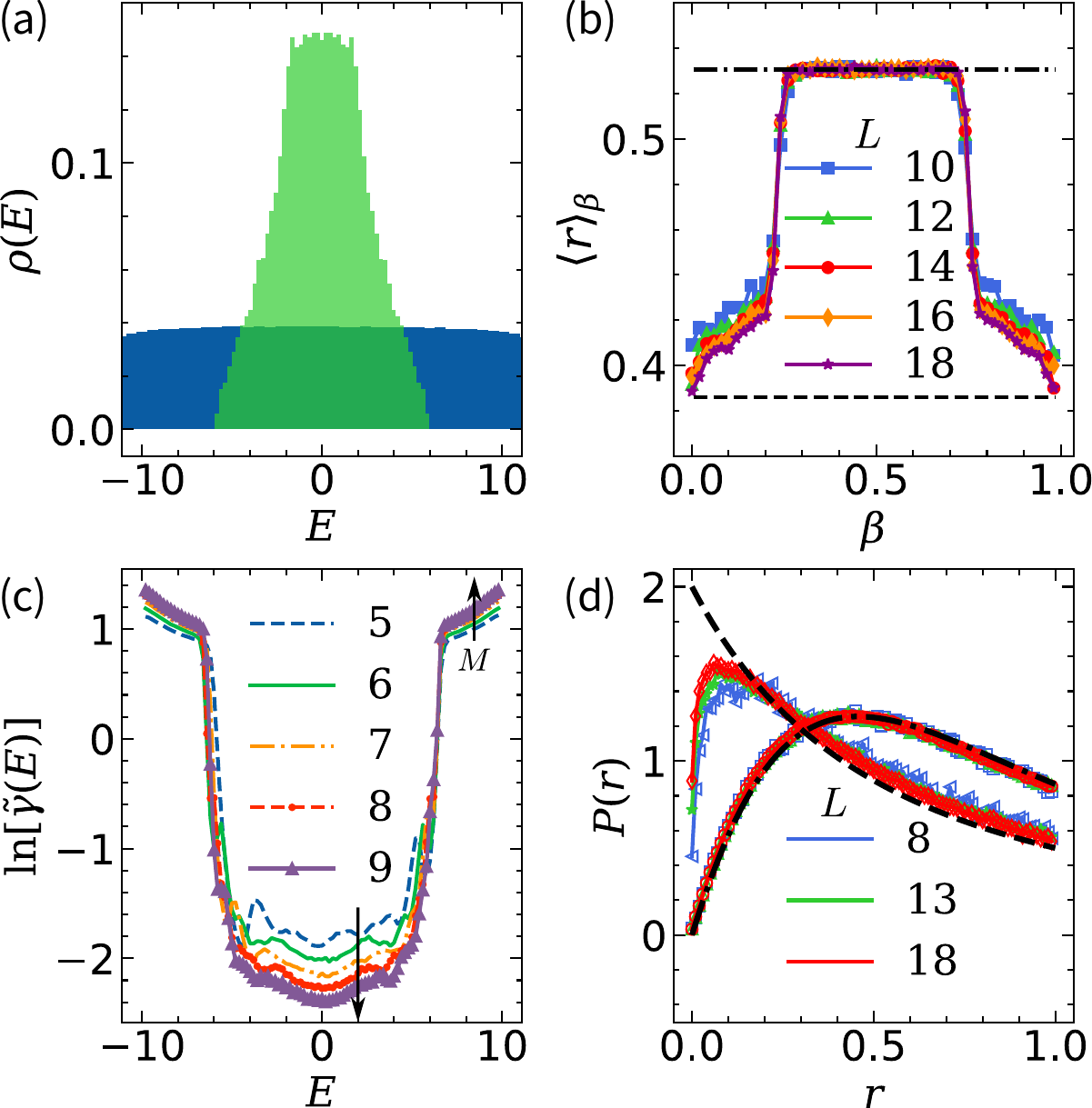}
\caption{
(a) The averaged density of states $\rho(E)$ of Eq. \ref{eq-al-model} in 3d, with $L=18$ (blue shadow). The green shadow is $\rho(E)$ for the 3d AL model with $V = 5$. (b) Level-spacing ratio $\langle r\rangle_\beta$ against energy with different sizes $L$ at $V_1=5$, $V_2=20$, $M_1=M_2=0$ and $t_\text{v}=0.5$.
(c) $\tilde{\gamma}(E)$ versus energy $E$ with parameters the same as (b).  
(d) $P(r)$ for $|\beta-0.5|<0.15$ for GOE and $\beta<0.2$ for PE. }
\label{fig3-3d-al}
\end{figure}

\begin{figure}[htbp]
\centering
\includegraphics[width=0.48\textwidth]{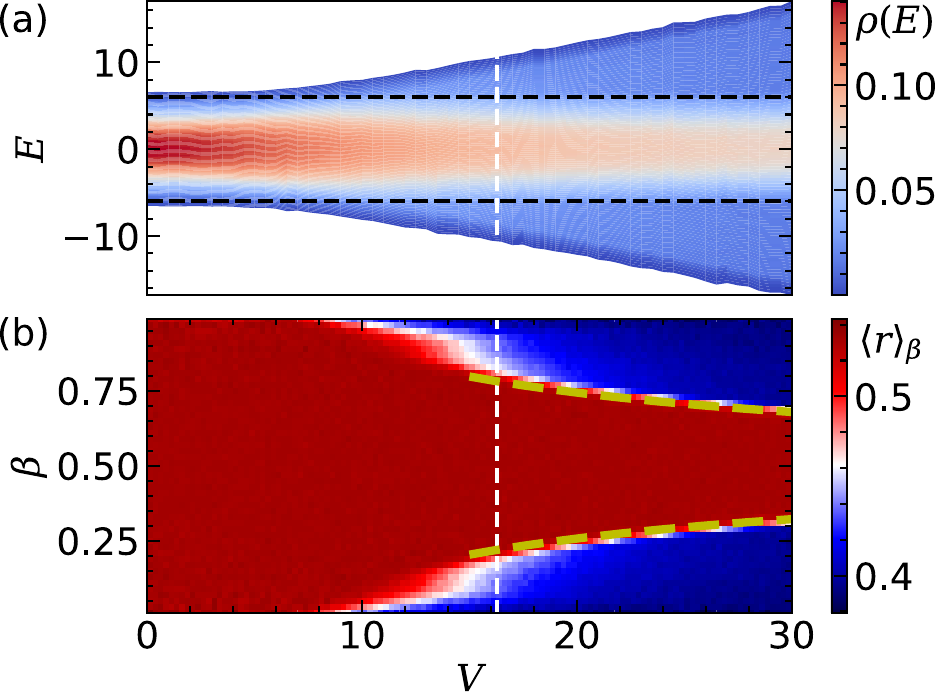}
\caption{
(a) $\rho(E)$ versus $V$ and $E$ without inter-component coupling, $t_\text{v} = 0$. The two black dashed lines denote the boundaries of the overlapped regime.
(b) $\langle r \rangle_{\beta}$ versus 
normalized energy $\beta$ and disorder strength $V$. Here we have used $L=18$ and averaged over $500$ realizations. MEs emerge when $V>V_c \simeq 16.3 t$ (white line) \cite{MacKinnon1981OneParameter}. 
The yellow lines represent the MEs in the overlapped spectra given by $\beta_c = 0.5 \pm \mathcal{A}/ V$, with $\mathcal{A} = 6.2 t$.}
\label{fig-ratio-large-disorder}
\end{figure}

\begin{figure}[htbp]
\centering
\includegraphics[width=0.48\textwidth]{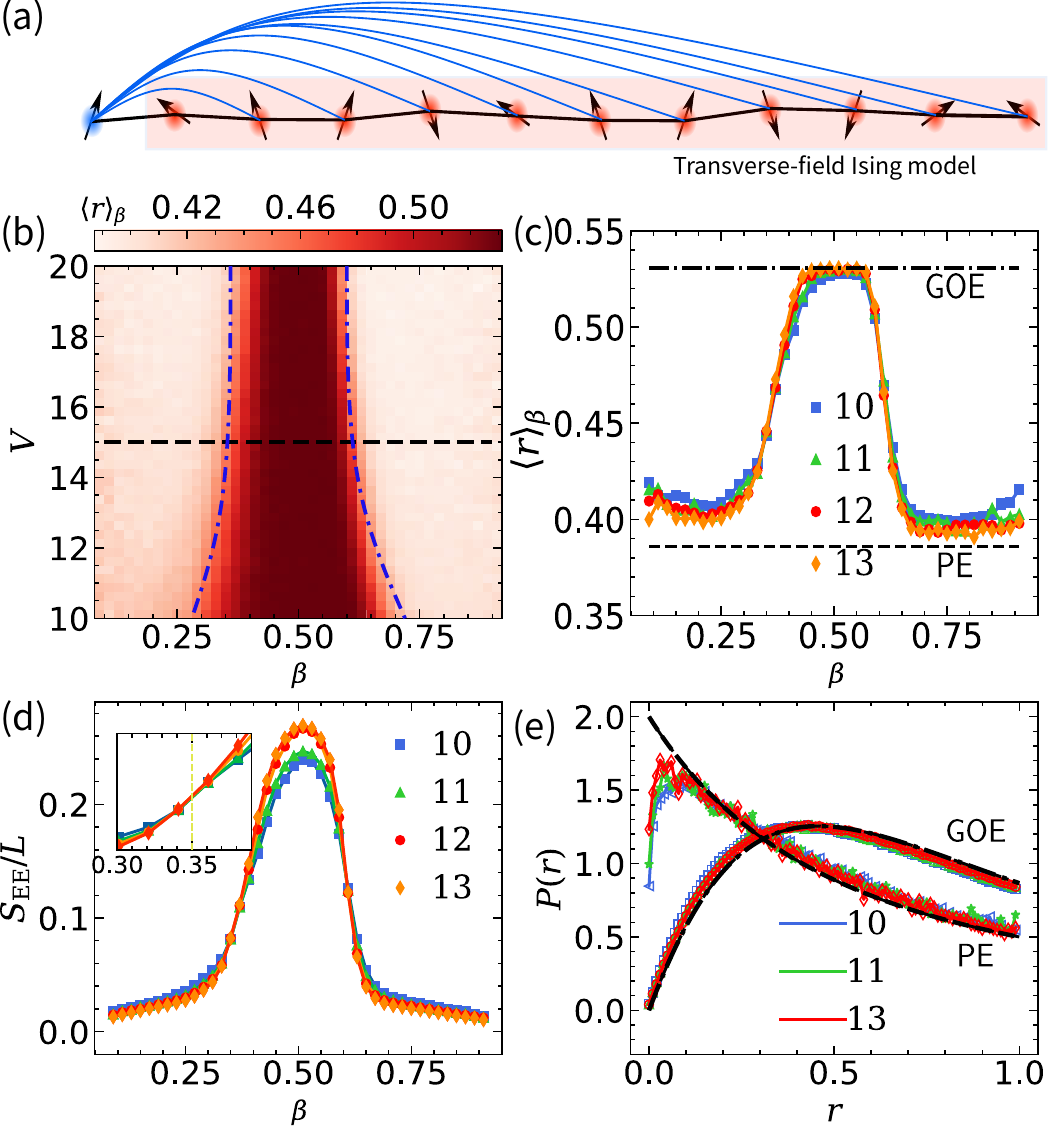}
\caption{
(a) Realization of MBL by coupling between a single spin and a transverse Ising model with long-range coupling. 
(b) ME in this model with $0.1<\beta<0.9$, $V_1=1$ and $L=13$. The blue dot-dashed lines are a guide for the eyes. 
(c) Detailed plot of ratio $\langle r \rangle_\beta$ at different sizes $L = 10 - 13$ with $V=15$; with $V_c \sim 9$.
(d) Normalized half-chain entanglement entropy $S_{\mathrm{EE}}/L$ against $\beta$ for $L=10 - 13$. The inset shows a detailed cross point (or ME) at $\beta_c = 0.35$.
(e) $P(r)$ with $|\beta-0.5|<0.1$ for GOE and $0.1<\beta<0.2$ for PE; see \onlinecite{SI}. }
\label{fig4-spin-mbl}
\end{figure}

\textit{(II) ME in 3d coupled model}: Disordered 3d models support AL at finite disorder strength \cite{abrahams_scaling_1979, anderson_absence_1958}.
In the spinless model, $V_c \sim 16.3t$ near $E \sim 0$ \citep{Hoffmanbook, Jung2012Finite, Alvermann2005Characterisation, Sierant2020Thouless}.
Thus, by choosing the disorder strength that $V_1<V_c$ and $V_2>V_c$, the inter-component coupling can induce direct coupling between localized and extended states, thus in the overlapped spectra, we expect the ME.
To this end, we fixed $V_1=5$, $t_\text{v} = 0.5$ and set $V_2 = V$ as a varying parameter.
We plot the ratio $\langle r\rangle_\beta$ against the normalized energy 
\begin{equation}
\beta = (E-E_\text{min})/(E_\text{max}-E_\text{min}),
\end{equation}
with different lattice sizes $L$ ($N = 2L^3$) in Fig. \ref{fig3-3d-al} (b).
The ratio $\langle r \rangle_\beta \rightarrow 0.5307$ in $|\beta-0.5|<0.15$ with the increasing of system size, demonstrating GOE statistics. Meanwhile, $\langle r \rangle_\beta \rightarrow 0.386$ in $\beta < 0.2$, demonstrating of Poisson statistics.
The distributions of $P(r)$ \citep{SI} in these regimes in Fig. \ref{fig3-3d-al} (d) also yield the same conclusion. Furthermore, we employ the transfer matrix method \citep{Tarquini2017Critical, Hoffmanbook} and study the dimensionless Lyapunov exponent $\tilde{\gamma}(E) = M \gamma(E)$ as a function of the energy $E$ for various cross-section length $M$.
We find that in the regime $|E| < 6$, $\tilde{\gamma}(E) \sim 0$ for the extended phase (see Fig. \ref{fig3-3d-al} (c)). 
However, outside this region, $\tilde{\gamma}(E)$ increases as a function of $M$ for the localized phase.  
This result is completely consistent with the level-spacing ratio from the eigenvalues. 
We also examine the density of state $\rho(E)$ and ratio $\langle r\rangle_E$ as a function of $V$, which are presented in Fig. \ref{fig-ratio-large-disorder}. From Fig. \ref{fig4-spin-mbl} (b), we expect the strong disorder to modify the boundaries in a way $\beta_c \sim 0.5 \pm  \mathcal{A}/V$, with $V > V_c$. This method can also be applied to disordered models with $d \ge 4$ \citep{Tarquini2017Critical, Jani2005Mean-field}. 

\textit{(III) ME in the coupled MBL model}: Finally, we verify the universality of our mechanism in the disordered many-body model. 
The construction of a many-body model is totally different from the single-particle model because the lattice site can not label the Hilbert space.
Namely, coupling two many-body chains can not generate a Hamiltonian in Eq. \ref{eq-couple-matrix}, but a Hamiltonian with much more complicated structures from the tensor product of two Hilbert spaces $\mathcal{K} = \mathcal{K}_1\otimes 
\mathcal{K}_2$.
To this end, we consider the following many-body model
\begin{eqnarray}
 \mathcal{H} && = | 0 \rangle\langle 0|\otimes H_\text{ext} + | 1 \rangle\langle 1|\otimes H_\text{loc}  + | 0 \rangle\langle 1| \otimes H_\text{c} + \mathrm{h.c.} \nonumber \\ 
            && = \mathbf{1}_{0}\otimes H_\text{id} + \sigma_0^z \otimes H_\text{diff} + \sigma_0^x \otimes H_\text{c},
\end{eqnarray}
with $H_\text{id} = (H_\text{ext} + H_\text{loc})/2$ and $H_\text{diff} = (H_\text{ext} - H_\text{loc})/2$.
This model can be regarded as a single spin coupled to the many-body model $H_\text{id}$.
The many-body localization transition has been well-established in the disordered spin chain model
\citep{Pal2010Many-body, Bera2015Many, Smith2016Many, Schreiber2015Observation}.

In the main text, we consider the following model
\begin{eqnarray}
	H_\text{a} = \sum_{m=1} J_{m}\sigma_m^x\sigma_{m+1}^x -J' \sigma_m^x \sigma _{m+2}^x+ \mathbf{B}_m^a \cdot \mathbf{\sigma}_m,
 \label{eq-mbl-spin}
\end{eqnarray}
where ${\bf B}_{m}^a = (h, 0, V_m^a)$ for $a = \text{ext}$ ($\uparrow$), loc ($\downarrow$), $J_m \in U(-1.05,-0.95)$, $V^{a}_m \in U(-V_{a}/2,V_{a}/2)$ and the coupling between the two components is given by 
\begin{equation}
    H_\text{c} = -g (\sigma_1^x + J' \sigma_2^x).
\end{equation}
Noticed that $H_\text{diff}$ induces long-range spin-spin couplings  $\sigma_0^z \sigma_m^z$ without spin-flip (see blue lines in Fig. \ref{fig4-spin-mbl} (a)). 
These long-range couplings are realizable in state-of-the-art digital simulation experiments with superconducting qubits \citep{Salathe2015Digital, Garcia2017Digital, Barends2015Digital, Xu2022Quantum}. 
When $g=0$, the Hilbert space is decoupled into two blocks, which correspond to $H_\text{ext}$ and $H_\text{loc}$ if we set $V_1 = 1 < V_c$ and $V_2 = V > V_c$. Then $H_\text{c}$ couples the two different components and satisfies the pedagogical model in Eq. \ref{eq-couple-matrix}. 
Here, we set $J'=0.3$, $h=0.6$, $g=1$ and $V_2 = V$. We find Eq. \ref{eq-mbl-spin} has MBL transition at $V_c \sim 9 $ (see S6 \citep{SI}). The level-spacing ratio versus the disorder strength at different energy regions, with ME, is presented in Fig. \ref{fig4-spin-mbl} (b). The finite size analysis up to $N = 2^{L+1} = 16384$ in Fig. \ref{fig4-spin-mbl} (c) and (d) also demonstrate that the overlapped spectra are ergodic, while the feature of the un-overlapped spectra is unchanged (or MBL). Furthermore, we find that in the MBL  (ergodic) phase, $S_\text{EE}/L$ 
decreases (increases) with increasing  system size, demonstrating that these two phases have different scaling laws in their entanglement entropy \citep{Bauer2013Area, Luitz2015Many-body, Decker2022Many-body}, with a boundary at $\beta_c \simeq 0.35$ for $V = 15$.
We have also presented in Fig. \ref{fig4-spin-mbl} (e) the level spacing distribution in these two phases, showing that in the ergodic (MBL) phase, $P(r)$ satisfies the expression of GOE (PE).
In S5 \citep{SI}, we have also considered a different realization of this disordered model based on $\mathbb{Z}_2$ symmetry for $H_\text{loc}$ and $H_\text{ext}$, and symmetry-breaking terms $H_\text{c}$, which exhibit the same conclusion, demonstrating the universality of our prediction from the model in Eq. \ref{eq-couple-matrix}.
We hope this finding can be applied to much wider systems for MBL and the associated MEs, and facilitate their experimental confirmation. 

\textit{Conclusion and discussion}: We present a unified mechanism for MEs in disordered single-particle and many-body models using coupled models, in which the position of MEs is adjustable on demand.
There are many intriguing directions motivated by this work.  
Firstly, it provides a generic way to construct MEs in physical models, offering exceptional opportunities for experimental verification of this long-sought controversial issue in MBL 
\citep{Deroeck2016absence, Deroeck2017Stability, Wei2019Investigating, Luitz2015Many-body, Brighi2020Stability, Deng2017Many-body, Chanda2020Many-body, Nag2017many-body, Modak2015Many-body,  Zhang2022Localization, Lazarides2015Fate}. 
Secondly, it could be applied to the possible MEs in non-Hermitian random models in the 
complex plane \citep{Tomasi2022Non-Hermitian, Hamazaki2019Non-hermitian}. 
Thirdly, it could be generalized to multi-block models, such as multiply coupled chains or many-body models, and even to much broader random ensembles \citep{Forrester2010LogGas}, for MEs. 
Finally, since the GOE can also be realized in some integrable models \citep{DAlessio2016Quantum, haake1991quantum}, this idea may even be used to realize the disorder-free ME.
Thus, this generic mechanism could greatly advance our understanding of ME and expand our knowledge of the mechanism of many-body delocalization \citep{Thiery2018Many, Rubio2019Many, See2022Many}. 

\textit{Acknowledgments}: 
This work is supported by the National Natural Science Foundation of China (NSFC) with No. 11774328, and Innovation Program for Quantum Science and Technology (No. 2021ZD0301200 and No. 2021ZD0301500).

\bibliography{ref} 
\end{document}


\preprint{APS/123-QED}

\title{Supplementary material: From single-particle to many-body mobility edges and the fate of overlapped spectra in coupled disorder models}

\author{Xiaoshui Lin}
\affiliation{CAS Key Laboratory of Quantum Information, University of Science and Technology of China, Hefei, 230026, China}
\author{Ming Gong}
\email{gongm@ustc.edu.cn}
\affiliation{CAS Key Laboratory of Quantum Information, University of Science and Technology of China, Hefei, 230026, China}
\affiliation{Synergetic Innovation Center of Quantum Information and Quantum Physics, University of Science and Technology of China, Hefei, Anhui 230026, China}
\affiliation{Hefei National Laboratory, University of Science and Technology of China, Hefei 230088, China}
\author{Guang-Can Guo}
\affiliation{CAS Key Laboratory of Quantum Information, University of Science and Technology of China, Hefei, 230026, China}
\affiliation{Synergetic Innovation Center of Quantum Information and Quantum Physics, University of Science and Technology of China, Hefei, Anhui 230026, China}
\affiliation{Hefei National Laboratory, University of Science and Technology of China, Hefei 230088, China}

\date{\today}

\maketitle

\tableofcontents
\clearpage

\section{Details about the numerical methods}

In this work, we use several complementary approaches to characterize 
the localization properties of wave functions in the coupled disorder system (see Eq. 1 in the main text). Some additional details about 
them are given below. 

\begin{enumerate}
\item {\bf Shift-invert method}

We employ the exact diagonalization method to obtain all the eigenvalues and eigenvectors when the matrix size is less than 
1.5$\times 10^4$. For the much larger matrices ($N > 2 \times 10^4$), 
we use the shift-invert method to obtain the eigenvalues and eigenvectors for a given energy $E$ based on sparse matrix technique. This method maps the eigenvalue $\lambda_i$ of the original Hamiltonian $H$ to the eigenvalue 
\begin{equation}
\tilde{\lambda}_i = (\lambda_i - E)^{-1}, \quad \text{of} \quad \tilde{\mathcal{H}} = \frac{1}{\mathcal{H}-E}.
\end{equation}
The eigenvalues $\lambda_i \sim E$ are obtained using the Lanczos algorithm. In the practical calculation, we make use of the sparse matrix, and the largest size we can 
calculate is around $2 \times 10^6$. We keep $N_E \sim 6-20$ eigenvalues $\lambda_i$ around $E$ and eigenstates $|\psi_{i}\rangle$ in each computation, averaged about $10^2 - 10^3$ realizations, which is sufficient to obtain a converged $\langle r\rangle_E$. 

\item {\bf Iterative transfer matrix method}

The equation of the tight-binding model can be written as 
\begin{equation}
    J \Psi_{n+1} + J \Psi_{n-1} + V_n \Psi_n = E\Psi_n,
\end{equation}
with $\Psi_n = [\psi_{1}, \psi_{2},\dots, \psi_{M^{d-1}}]^T$ is the vector of the wave function in the $n$-th cross-section. We have the following iteration equation based on the transfer matrix as 
\begin{equation}
  \begin{pmatrix}
      \Psi_{L+1} \\
      \Psi_{L}
  \end{pmatrix} = \begin{pmatrix}
      J^{-1}(E - V_L) & 1 \\
      -1 & 0\end{pmatrix} \begin{pmatrix}
      \Psi_{L} \\
      \Psi_{L-1}
  \end{pmatrix} = T_L \begin{pmatrix}
      \Psi_{L} \\
      \Psi_{L-1}
  \end{pmatrix} = T_L T_{L-1} \cdots T_{1} \begin{pmatrix}
      \Psi_{1} \\
      \Psi_{0}
  \end{pmatrix} = \mathcal{T}_L\begin{pmatrix}
      \Psi_{1} \\
      \Psi_{0}
  \end{pmatrix}.
\end{equation}
Then the Oseledets multiplicative ergodic theorem guarantees the 
existence of the following limit  
\begin{equation}
\Gamma(E) = \lim_{L\rightarrow\infty}\frac{1}{2L}\ln(\mathcal{T}^{\dagger}_L\mathcal{T}_L),    
\end{equation}
and the Lyapunov exponent $\gamma(E)$ is defined by the smallest positive eigenvalue of $\Gamma (E)$. In the numerical calculation, we apply iterative QR decomposition \citep{Hoffmanbook} on each transfer matrix and perform the iterative step to a large size ($L \sim 10^6$). 
The system we used in the numerical calculation is actually a quasi-1d system with cross-section length $M \ll L$.
Thus, to account for the finite-size effect, we need to define the dimensionless Lyapunov exponent $\tilde{\gamma}(E) = M \gamma(E)$, which reads as \citep{Tarquini2017Critical}
\begin{equation}
    \tilde{\gamma}(E) \sim \left\{ \begin{array}{cc}
       (M \gamma(E))   &  \text{localized,} \\
       \text{constant} &  \text{critical,}  \\
       (\gamma(E)/M)^{d-2}  &  \text{extended.}  \\
    \end{array}\right.
\end{equation}
Thus, this criterion is very useful for the $d\geq 3$ AL models and the associated ME. 
\end{enumerate}
 
\section{Energy shift for tunable MEs}

\begin{figure}[!htbp]
\centering
\includegraphics[width=0.7\textwidth]{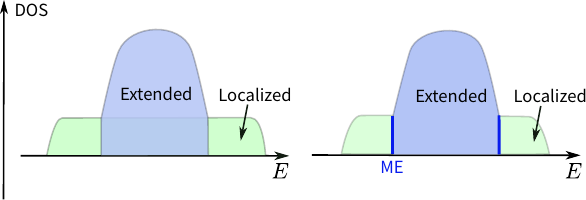}
\caption{Tunable MEs in coupled disordered models. Here, the extended states are located in the center of the localized states, giving rise to 
two MEs in the presence of inter-component coupling. In the un-overlapped 
spectra, all states are still localized. The position of the MEs can be
tuned by the energy shift $M_\alpha$. }
\label{supfig-schematic}
\end{figure}

One of the major advantages of our scheme for MEs is its tunability in 
experiments by the energy shift $M_\alpha$. In the main text, we consider the possible ME for the overlap structure in Fig. 1 (c) with only one ME for models with $d\leq 2$ and two MEs for the 3d model; 
and here, we can tune the position of the overlapped spectra, yielding two MEs for $d\leq 2$ models and one ME for the 3d model on demand.
The mechanism is presented in Fig. \ref{supfig-schematic}. In this section, we consider the energy shift of $M_1$ for $H_\text{ext}$, and we consider the physics in 1d, 2d models with $M_1 = 0$ and 3d models with $M_1 = -15$.  

\begin{figure}[!htbp]
    \centering
\includegraphics[width=0.75\textwidth]{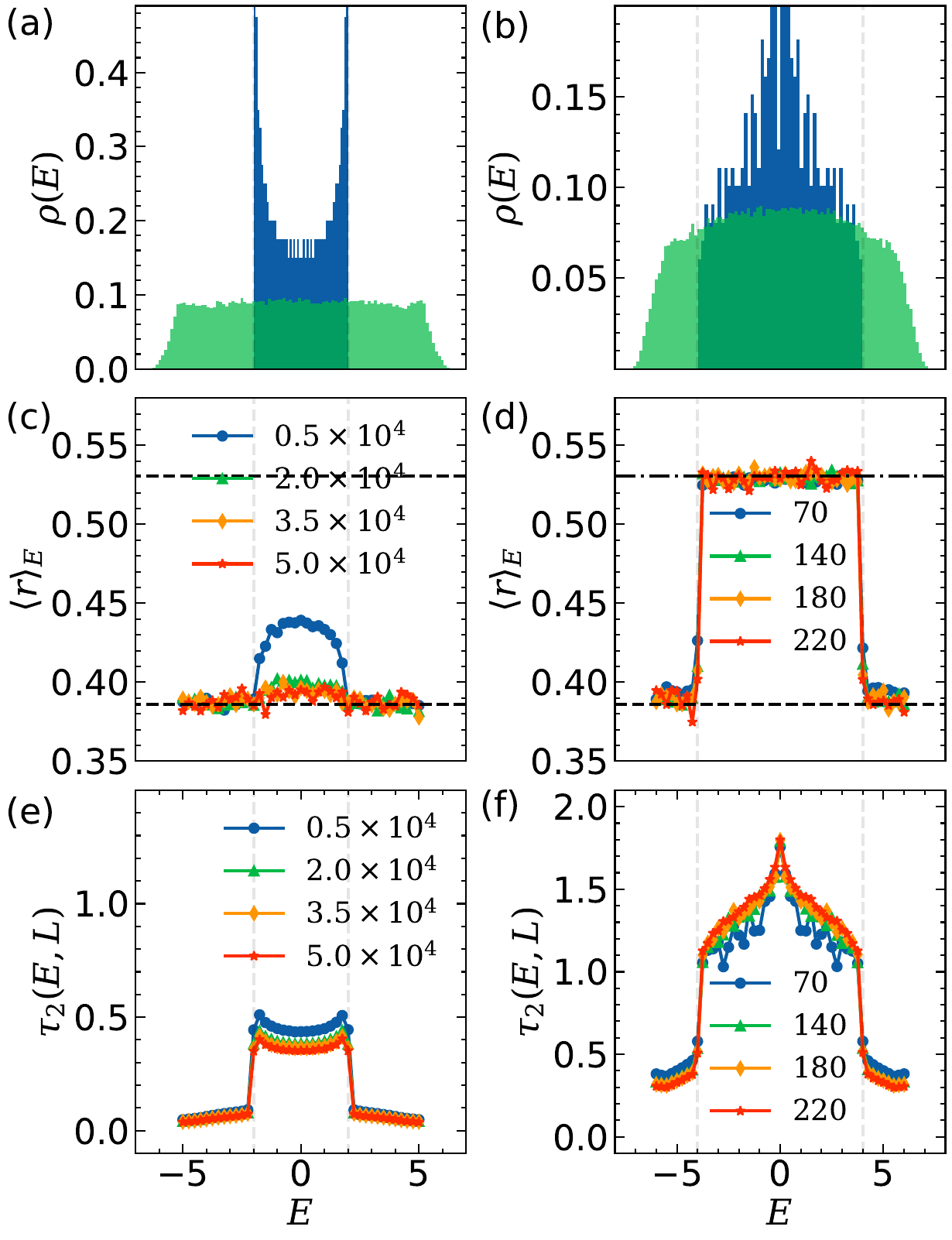}
    \caption{The average density of states, $\rho(E)$, of the uncoupled ($t_\text{v} = 0$) models 
    in 1d (a) and 2d (b) models, in which the blue and green shadow represents $\rho(E)$ of 
    $H_\text{ext}$ ($H_\text{loc}$), respectively. (c) and 
    (e) the level-spacing ratio $\langle r\rangle_E$ and the fractal dimension $\tau_2(E, L)$ 
    of the 1d coupled disorder model for sizes $L = (0.5 - 5.0) \times 10^4$. 
    (d) and (f) are the corresponding results in 2d coupled disordered model with $L = 70 - 220$ ($N=2L^2$). The vertical grey dashed lines denote the boundaries of the overlapped spectra. 
    Here we have used $V_1=M_1 = M_2=0$, $V_2=10$, $t_\text{v}=0.1$ in all figures.}
    \label{supfig-1d-2d}
\end{figure}

\begin{figure}[!htbp]
    \centering
\includegraphics[width=0.75\textwidth]{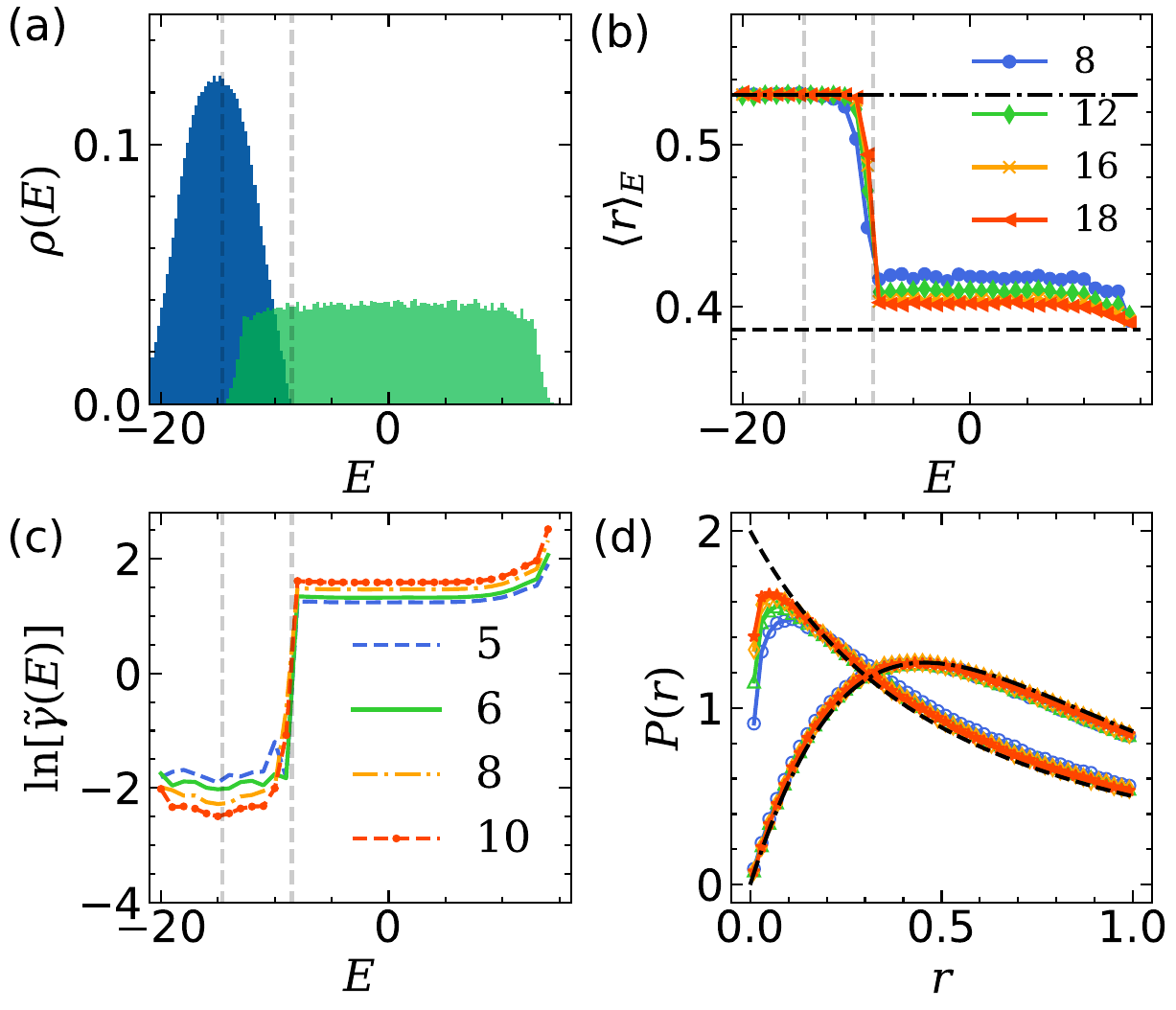}
\caption{(a) The average density of states, $\rho(E)$, of the uncoupled ($t_\text{v} = 0$) 3d disordered model, in which the blue and green shadow represents $\rho(E)$ of 
$H_\text{ext}$ ($H_\text{loc}$), respectively. 
The vertical grey dashed lines represent the boundaries of the overlapped regime (with $E=-14.6$ and $E=-8.5$). 
(b) The level-spacing ratio $\langle r\rangle_E$ for this 3d disordered model, in which the two horizon dashed lines denote the value of PE and GOE.
(c) The dimensionless Lyapunov exponent $\tilde{\gamma}(E)$ versus the energy for different cross-section length $M$ in this 3d model, with iteration step up to $L = 10^6$ and $N = M^2 L$. 
(d) The level-spacing ratio distribution $P(r)$ for energy levels with $E<-8.5$ ($E>-8.5$) approaches GOE (PE) when the system size is large enough. In (b) - 
(d), we have used $V_1 = 5$, $M_2 = 0$, $V_2 = 25$, and $t_\text{v} = 0.1$. }
\label{supfig-3d}
\end{figure}

\begin{enumerate}
\item The results for 1d and 2d disordered models are shown in Fig. \ref{supfig-1d-2d}. We use $V_1 = M_1 = M_2 = 0$, $V_2 = 10$, and $t_\text{v} = 0.1$ for all figures, where the overlapped regime is always located in the center of the spectra. In the 1d system, all states are localized with $\langle r\rangle_{E} \sim 0.386$ for all $E$. However, in the 2d disordered models, we find that in the overlapped spectra, $\langle r\rangle_{E} \sim 0.53$, and in the un-overlapped spectra, $\langle r\rangle_{E} \sim 0.386$. These results indicate that while all
states are localized in the 1d disordered models, extended phase can be
realized in the 2d disordered models using our approach. In the 2d models, 
one should realize the tunable MEs. While these models have two MEs, the major conclusions are the same as those presented in the main text. 

\item The results for the 3d  model with $M_1=-15$, with only one ME, are presented in Fig. \ref{supfig-3d}, in which we find that all the states in the overlapped regime are extended with $\langle r \rangle_E \sim 0.53$. The states in the un-overlapped regimes are extended( $E<-14.6$) with $\langle r \rangle_E \sim 0.53$ and localized ($E>-8.5$) with $\langle r \rangle_E \sim 0.386$, thus the ME at $E \simeq -8.5$. This can also be viewed from the dimensionless Lyapunov exponent $\tilde{\gamma}(E)$ presented in Fig. \ref{supfig-3d} (c), in which when $E > - 8.5$, $\tilde{\gamma}(E)$ is finite, yet when $E < -8.5$, $\tilde{\gamma}(E)$ approaches zero. 
In Fig. \ref{supfig-3d} (d), the level-spacing ratio distribution is presemted, which exhibits the same signature. 
\end{enumerate} 

\section{Finite size scaling of the fractal dimension}

\begin{figure}[!htbp]
    \centering
\includegraphics[width=0.75\textwidth]{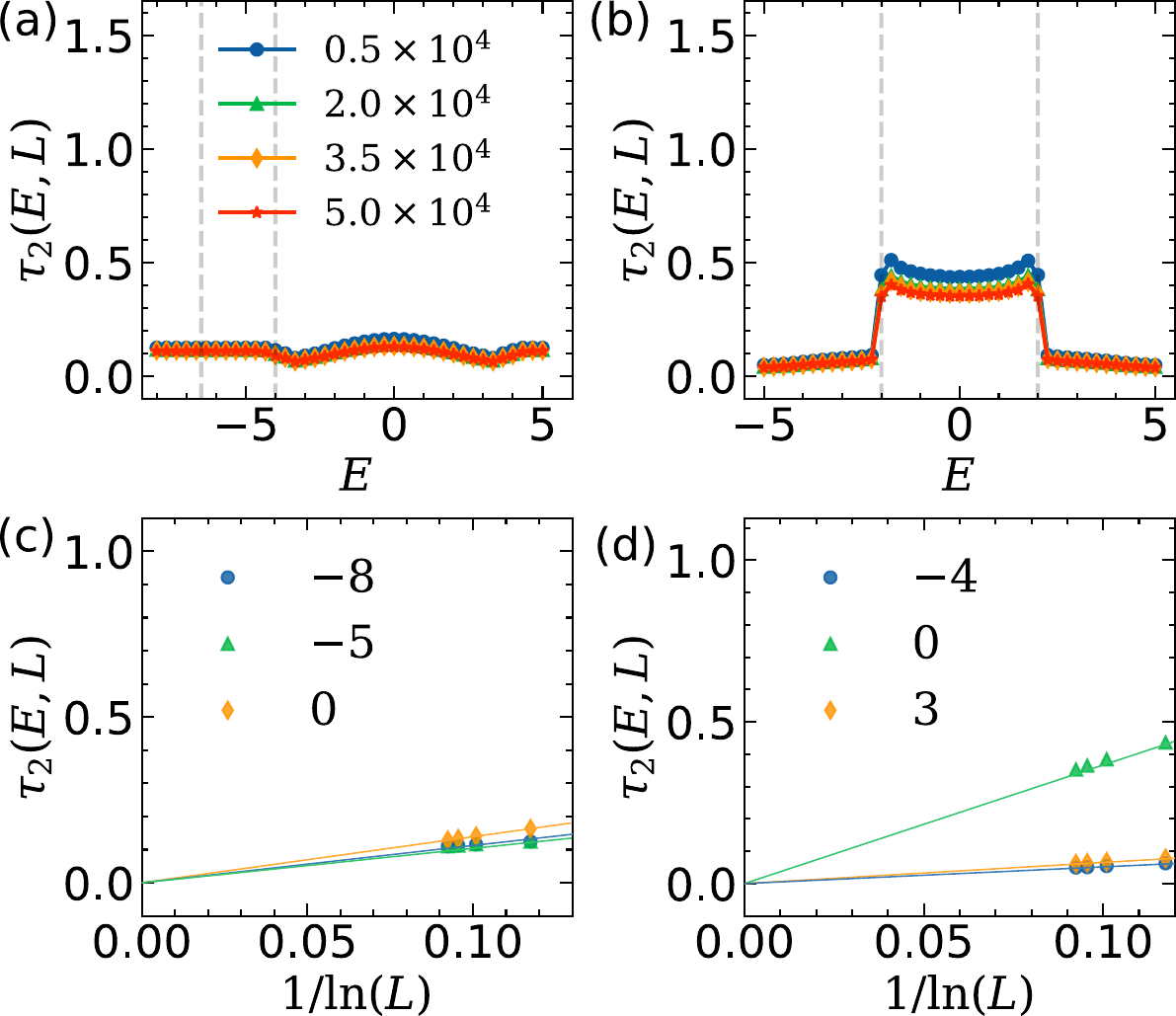}
\caption{The fractal dimension of the 1d disordered models, with energy shift (a) $M_1=-6$ and (b) $M_1=0$. (c) and (d) are the corresponding scaling of fractional dimension as a function of $1/\ln(L)$, with $L = 5\times 10^3$ to $L= 5 \times 10^4$. The different points represent the data for different energies $E$. The solid lines are a guide for the eyes.}
\label{supfig-1d-scale}
\end{figure}

\begin{figure}[!htbp]
    \centering
\includegraphics[width=0.75\textwidth]{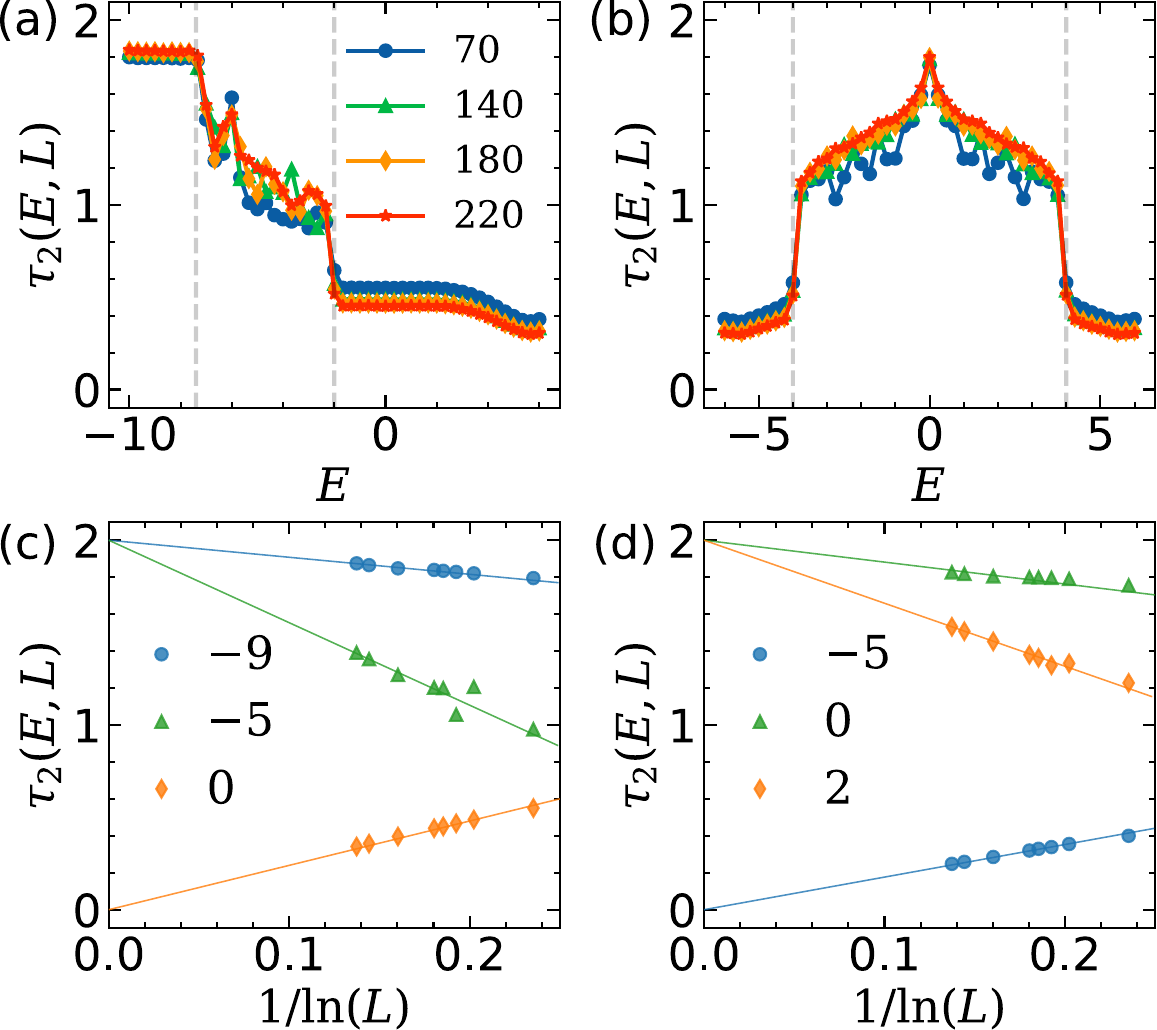}
\caption{The fractal dimension of the 2d disordered models, with energy shift (a) $M_1=-6$ and (b) $M_1=0$. (c) and (d) are the corresponding scaling of fractional dimension as a function of $1/\ln(L)$, with the size ranging from $L=70$ to $L=1450$, where the total number of sites is $N = 2L^2$.
The different points represent the 
data for different energies $E$.
The solid lines are a guide for the eyes.}
\label{supfig-2d-scale}
\end{figure}

The fractal dimension $\tau_2(E, L)$ can be used to distinguish the extended
states and localized states, and even critical states. Here, we focus on the finite scaling of fractal dimensions in 1d and 2d systems. By definition, we should have 
\begin{equation}
    \tau_2(E, L) = \tau_2(E) + C/\ln(L), 
\end{equation}
where $C$ is a non-universal constant, and $\tau_2(E)$ is the thermodynamic limit 
of the fractional dimension $\tau_2(E, L)$. The results for the 1d and 2d coupled disordered models 
are presented in Fig. \ref{supfig-1d-2d} (e) and (f), Fig. \ref{supfig-1d-scale} (a) and (b), and Fig. \ref{supfig-2d-scale} (a) and (b).
The scaling of $\tau_2(E, L)$ as a function of $1/\ln(L)$ presented in 
Fig. \ref{supfig-1d-scale} (c) and (d) indicate that 
$\tau_2(E) \rightarrow 0$ for all $E$ in the thermodynamic limit.
Thus in the 1d models, all states are localized in the presence of coupling between the two components; 
see the picture of Fig. 1 (b) in the main text.
However, in the 2d disordered models, as shown in Fig. \ref{supfig-2d-scale} (c) and (d), we find $\tau_2(E) \rightarrow 2$ in the overlapped spectra of the coupled model with different $M_1$.
however, in the un-overlapped spectra, $\tau_2(E) \rightarrow 0$ for the localized phase. This result is consistent with the main text's major conclusions. 

\section{The level-spacing ratio distribution for 2d disordered model}

\begin{figure}[!htbp]
    \centering
\includegraphics[width=0.75\textwidth]{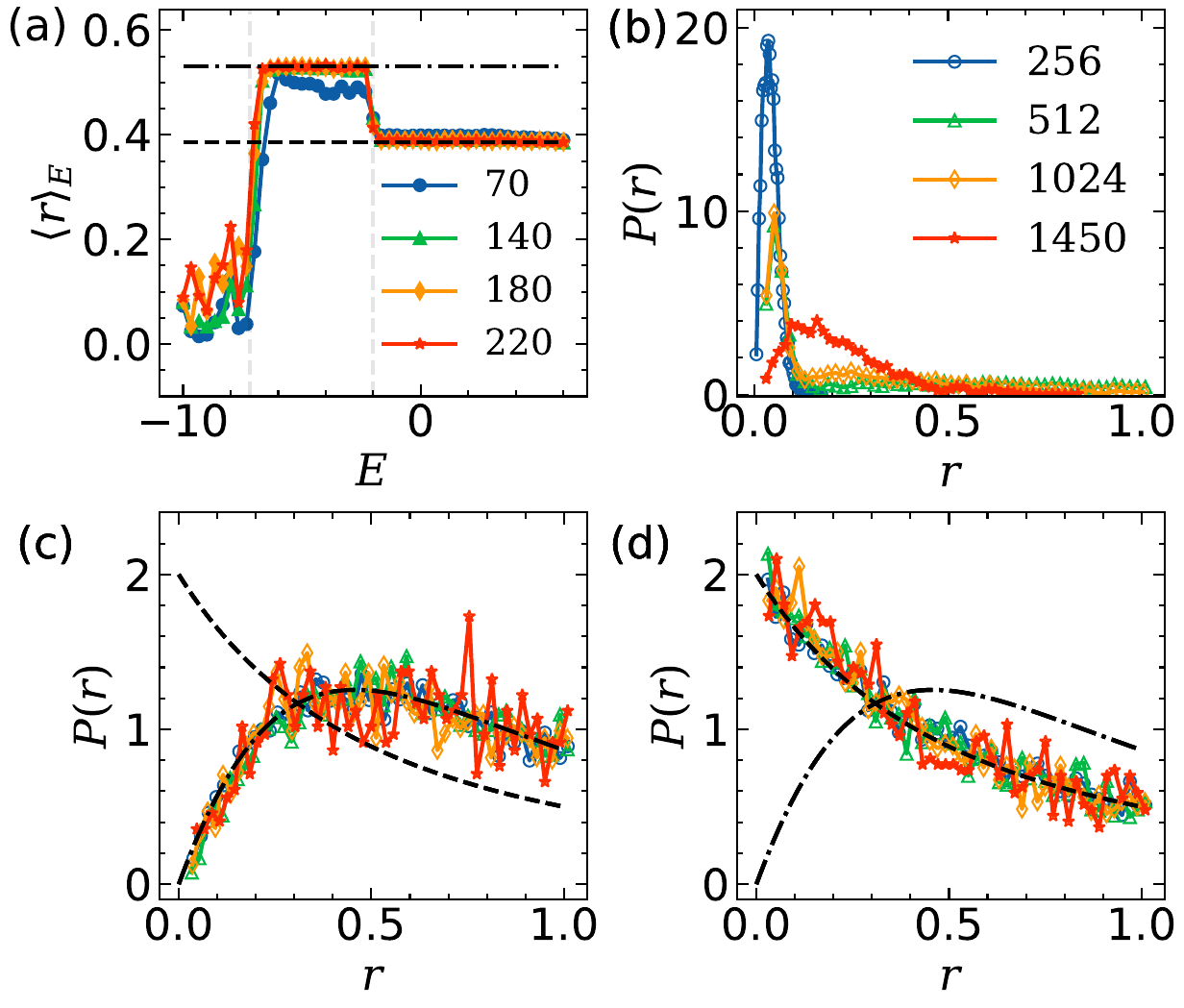}
\caption{The averaged level-spacing ratio $\langle r \rangle_{E}$ for different $E$ in 2d disordered model. (b) and 
(d) are the distribution of the level-spacing ratio $r$ for states with energy around (b) $E=-9$, (c) $E=-5$, and (d) $E=0$. The black dashed and dashed-dotted lines represent the 
expressions in Eq. \ref{eq-pr-pe} (for PE) and Eq. \ref{eq-pr-goe} (for GOE).}
\label{supfig-ratio-2d}
\end{figure}

Here, we present the results of the level-spacing ratio for the 2d disordered model, in which the parameters are chosen the same as that used in Fig. 2 (d) in the main text.
It is well-known fromt the random matrix theory that the distribution of the level-spacing ratio is approximately given by the Wigner-surmise \citep{Atas2013Distribution}. The closed expression for the Wigner-Dyson ensemble is given by 
\begin{equation}
P_{\beta}(r) = \frac{1}{Z_{\beta}} \frac{(r + r^2)^{\beta}}{(1 + r + r^2)^{1 + 3\beta/2}},
\label{eq-pr-goe}
\end{equation}
with $\beta$ the Dyson index for Guassian orthogonal ensemble ($\beta=1$), Guassian unitary ensemble ($\beta=2$), Guassian symplectic ensemble ($\beta=4$); and $Z_\beta$ is the corresponding normalization constant. The exact expression 
for the Poisson ensembles is given by 
\begin{equation}
P_\text{PE}(r) =  \frac{2}{(1 + r)^{2}}.
\label{eq-pr-pe}
\end{equation}
Here, the definition of $r$ can be found in the main text, with $r \in [0, 1]$.
These two distributions have been used in Fig. 3 (d) in the main text, and more details are presented in Fig. \ref{supfig-ratio-2d}.
While this distribution agrees well with the GOE around $E = -5$, and with PE around $E = 0$, the averaged $\langle r\rangle_E < 0.2$ can not be understood from these two distributions when the size of the system is less than $1450$; see $E = -9$ in Fig. \ref{supfig-ratio-2d} (a) and (b).
Here the largest system presented here is $L=1450$ with matrix size to be $N = 2L^2 = 4205000$.
We find that $P(r)$ is significantly different from the results in GOE or PE.
However, the value of $P(0)$ still decreases with the increasing of system size, indicating strong strong repulsion interactions between the energy levels, which is a typical feature of the extended phase. 
In combination with the fractal dimension in Fig. \ref{supfig-2d-scale} (c) and (d), we can conclude that states in this regime are extended. 

\begin{figure}[!htbp]
    \centering
\includegraphics[width=0.75\textwidth]{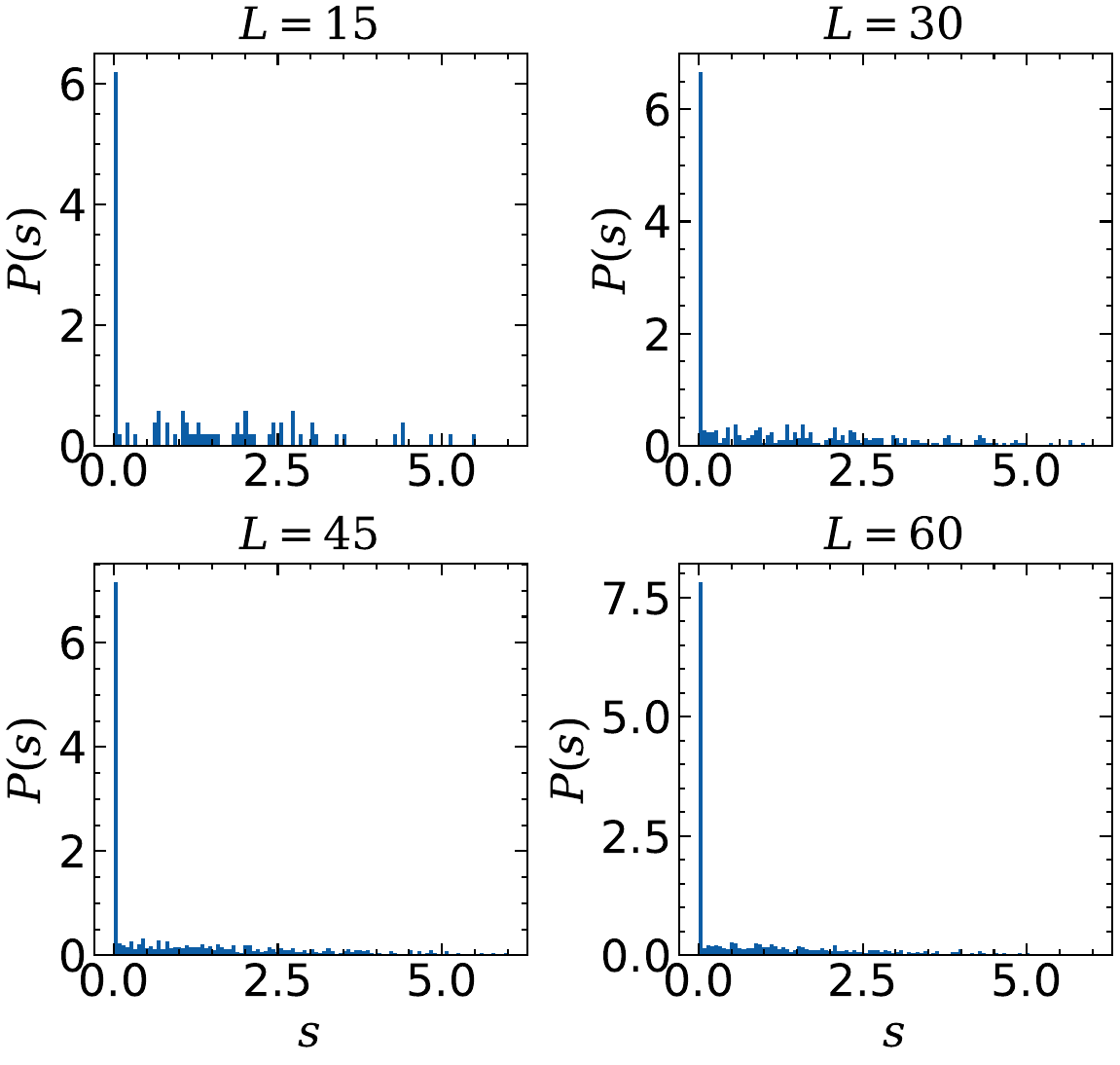}
\caption{The level-spacing distribution $P(s)$ in a clear Hamiltonian with different lattice sizes. The peak at $s = 0$ suggests energy-level degeneracy in higher dimensions ($d \ge 2$). } 
\label{supfig-2d-level-spacing}
\end{figure}

The repulsive interaction between the energy levels in the un-overlapped regime( see Fig. \ref{supfig-ratio-2d} (b)) arises from the inter-component coupling, otherwise, the degeneracy in the higher-dimensional ($d \ge 2$) models will lead to dramatically different distributions.
To understand this point, with $\langle r \rangle_E < 0.2$, we then consider the level-spacing distribution $P(s)$ for the model with Hamiltonian $H= \sum_{\langle i,j\rangle} a_i^{\dagger} a_j$ on a square 2d lattice.
In this clear Hamiltonian, all states are extended.
The results are presented in Fig. \ref{supfig-2d-level-spacing} for various system sizes $L$, with $N = L^2$, with $s_n = E_n - E_{n-1}$. The degeneracy becomes more and more significant with the increasing system size $L$. 
Thus in the presence of inter-component coupling, we find that the degeneracy at $E$ is lifted and the repulsive interaction between energy levels in the un-overlapped spectra becomes more and more important with increasing system size.
In this case, the states are still extended, yet an increase of $\tau_2(E, L)$ with the increase of $L$ is evident; see Fig. \ref{supfig-2d-scale} (a) and (b).
In this case, it's expected that, in the thermodynamic limit, the distribution $P(r)$ will approach GOE; see $L = 1450$ in Fig. \ref{supfig-ratio-2d} (b). 

\section{Many-body localizations and MEs in coupled disordered model}

\begin{figure}[htbp]
\centering
\includegraphics[width=0.75\textwidth]{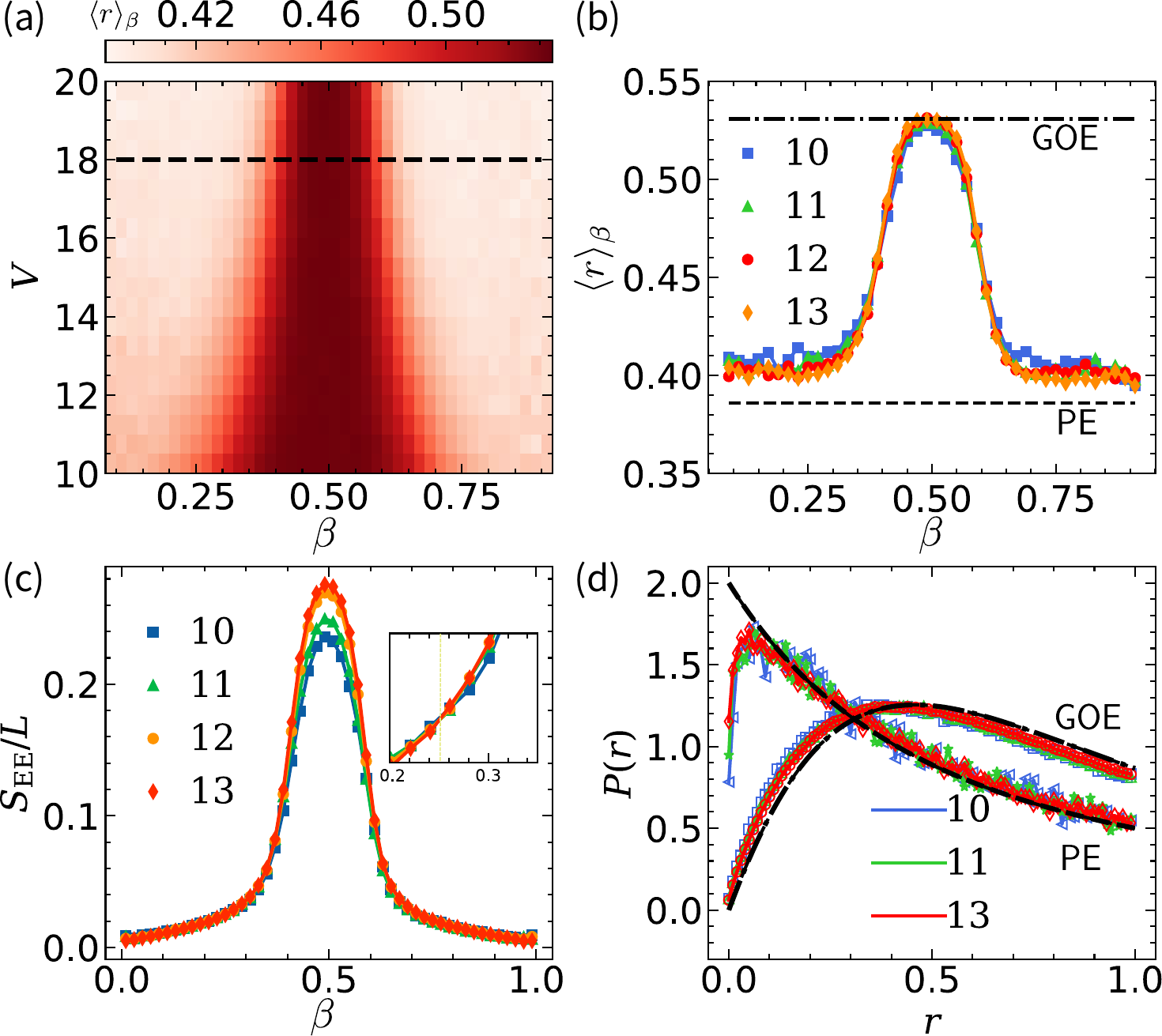}
\caption{
(a). The level-spacing ratio of the model \ref{eq-spin-symmetry}. The Red region donates the ergodic states while the white region denotes MBL states. (b).  The ratio $\langle r\rangle_\beta$ against energy with different size $L$ at $V=18$. (c). The normalized half-chain entanglement entropy $S_{\mathrm{EE}}/L$ against normalized energy windows $\beta$. The inset panel shows a detail of the cross point around $0.25$ of the lines. (d). The distribution $P(r)$ at different energy windows. The middle spectrum $|\beta-0.5|<0.05$ approaches GOE. The spectrum with $\beta <0.2$ approaches GOE. 
}
\label{supfig-spin-mbl-symmetry}
\end{figure}

We need a general way to construct the block matrix of Eq. 1 in the main text. In the disordered single-particle models, the Hilbert spaces can be well labeled by the lattice sites and internal spin degrees of freedom, thus the construction of the block matrix is straightforward.
In the many-body models, it will become much more complicated due to the tensor product of two Hilbert spaces. 
Here we show that it can be realized using some symmetry dependent disordered.
This is because symmetry is the most natural way to divide the whole Hamiltonian into block structures. 
Thus, the coupling between the blocks can be realized using some symmetry-breaking terms.
To this end, we consider the following 
many-body model
\begin{equation}
\begin{aligned}
H &= \sum_{i=0}^{L}\left( J_1 \sigma_i^x \sigma_{i+1}^x + J_2 \sigma_i^x \sigma_{i+2}^x \right)  + \sum_{i=1}^{L} V_{i}^{(1)}\frac{1+Z}{2}\sigma_{i}^z\frac{1+Z}{2} + V_{i}^{(2)}\frac{1-Z}{2}\sigma_{i}^z \frac{1-Z}{2} + h\sigma_i^{x}.
 \end{aligned}
 \label{eq-spin-symmetry}
\end{equation}
where the symmetry term is given by 
\begin{equation}
Z = \prod_i \sigma_i^z = \mathbb{Z}_2.
\end{equation}
Obviously, when $h = 0$, the Hamiltonian commute with $Z$, and the whole Hilbert space of $H$ can be divided into two blocks with $Z = +1$ and $Z = -1$. We denote these two Hamiltonians as $H_+$ and 
$H_-$, respectively, which are controlled by two independent random
numbers, $V_i^{(1)}$ and $V_i^{(2)}$. When this disorder strength is weak enough, the Hamiltonian belongs to the extended phase in GOE, that is, $H_\text{ext}$; and when the disorder strength is strong enough, it belongs to the MBL phase in PE, that is, $H_\text{loc}$. 
The symmetry breaking term of $h$ induces coupling between these two Hilbert spaces, thus plays the same role as $H_c$. This method provides
a general recipe to construct various many-body models, which can be described by the block matrix of Eq. 1 in the main text. The results for this model is presented in Fig. \ref{supfig-spin-mbl-symmetry}, in which all the basic features are the same as that in Fig. 5 in the main text, based on a different many-body disordered models. 

\section{The transition point of disordered transverse Ising model}

\begin{figure}[htbp]
\centering
\includegraphics[width=0.75\textwidth]{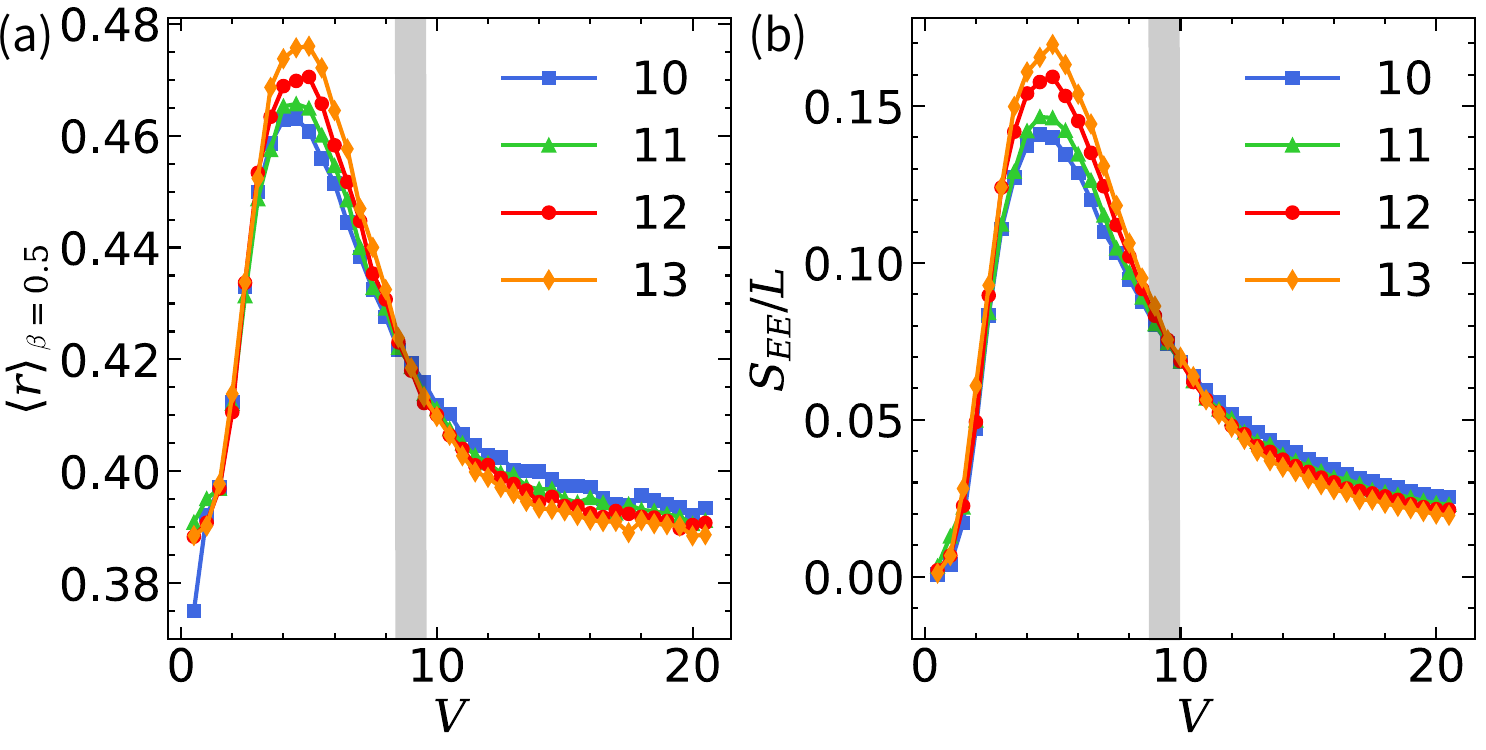}
\caption{
(a). The scaling analysis of the level-spacing ratio $\langle r\rangle_{\beta=0.5}$ and normalized entangle entropy $S_{EE}/L$ in the disordered many-body model studied in the main text. We find there is a critical point around $V_c \simeq 9$. The system hosts the MBL phase when $V > V_c$ and the ergodic phase when $V < V_c$. The model is integrable in the limit $V = 0$, and we find $\langle r\rangle_{\beta=0.5} \sim 0.4$.
}
\label{supfig-tfi-scale}
\end{figure}

It is important to verify the transition point of the MBL transition of the Ising model used in Fig. 5 in the main text. We have the following Hamiltonian,
\begin{equation}
	H = \sum_{m}  J_{m} \sigma_m^x\sigma_{m+1}^x - J' \sigma_m^x \sigma _{m+2}^x + h \sigma_m^x + V_m \sigma_m^z, 
\end{equation}
with $V_m \in U(-V/2,V/2)$ and $J_m \in U(-1.05,0.95)$. We set $h=0.6$, $J'=0.3$, the same as that used in the main text.
This model is in the MBL phase with strong disorders and in the
ergodic phase with weak disorders. We use the normalized entanglement entropy and mean level-spacing ratio to characterize this phase transition, where only states in the middle spectra $|\beta-0.5|<0.005$ were considered. 
The results are presented in Fig. \ref{supfig-tfi-scale}.
We find a critical point at $V_c \sim 9$. When $V < V_c \sim 9$, 
$\langle r\rangle_E$ and $S_\text{EE}/L$ increase with the increasing
of system size, indicating of ergodic phase. However, when s
$V > V_c$, these two quantities will decrease with the increasing of system size, indicating of MBL phase. This transition strength is used in Fig. 5 in the main text. 

\clearpage

\bibliography{ref}